\documentclass{article}
\usepackage{amsmath,amsfonts,amssymb,amsthm,latexsym,stmaryrd}
\usepackage[dvips]{graphicx,epsfig}
\usepackage[left=2.5cm,right=2.5cm,bottom=2.5cm,top=2.5cm]{geometry}

\begin{document}

\markboth{Soto-Eguibar  and Moya-Cessa} {Perturbative approach to diatomic lattices}

\title{Perturbative approach to diatomic lattices}

\author{Francisco Soto-Eguibar and H\'{e}ctor Manuel Moya-Cessa\\
        \small Instituto Nacional de Astrof\'{\i}sica, \'Optica y Electr\'onica (INAOE)\\
        \small Calle Luis Enrique Erro 1, Santa Mar\'{\i}a Tonantzintla, Puebla, 72810 Mexico
        \\
        \\
        Published as: Perturbative approach to diatomic lattices\\
        International Journal of Quantum Information\\
        Vol. 10, No. 6 (2012) 1250072 (9 pages)\\
        DOI: 10.1142/S0219749912500724}

\maketitle

\begin{abstract}
By using a small rotation approach, we show that it is possible to obtain well behaved perturbed solutions for the amplitude of the electromagnetic field propagating in a photonic waveguide array. This array mimics the propagation of a quasiparticle among the sites of an infinite one-dimensional chain.
\end{abstract}

\section{Introduction}
The analogy between linear lattices and  quantum mechanical interactions \cite{Crisp}  has been a fundamental step for the emulation, via classical systems, of quantum mechanical interactions such as the atom-field interaction, and, as a consequence ion-laser interactions \cite{Jona,Champi}. This is not only important  because of pure scientific reasons but also because of the possible applications in quantum information processing. In this latter case, the properties of classical systems have been used to realize quantum computational operations by quantum-like systems and, in particular, it has been show how a controlled-NOT gate may be generated in nonhomogeneous optical fibers \cite{Manko1}.
At a fundamental level, \textit{e.g.}, it has been possible to the emulate the most basic atom-field interaction, the Jaynes-Cummings model, theoretically \cite{Longhi} and experimentally \cite{Longhi2} with arrays of photonic waveguides; and, just to give another example, it has been proposed to model non-linear coherent states \cite{Manko2} in waveguide arrays \cite{Roberto}. Linear coherent states have also been modeled via linear arrays of photonic waveguides \cite{Keil}. \\
The analysis we produce here is twofold. On one hand we produce an analogy of a quantum system by using a classical one, while, on the other hand, we introduce concepts and techniques not know to the classical optics community that may serve them as powerful tools for their analysis of non-integrable problems.

\section{Diatomic waveguide arrays}
In a  diatomic photonic waveguide the field amplitude in each way is obtained form the  infinite system of coupled ordinary first order differential equations \cite{Longhi}
\begin{equation}\label{450100}
i\frac{du_n}{dz} = \omega (-1)^{n} u_n + \alpha (u_{n+1}+u_{n-1}), \qquad n=-\infty,...,\infty,
\end{equation}
where $\alpha$ and $\omega$ are arbitrary constants. The field at $z=0$ may be given, in general by
$u_n(z=0)=\psi_n$, for which we only ask to be normalized. \\
By itself, the solution of the problem represented by the system of equations (\ref{450100}) is interesting, but it will be shown as well that this array can mimic the propagation of a quasiparticle among the sites of an infinite one-dimensional chain. In order to solve it, the techniques of quantum optics are utilized; we can associate with the system (\ref{450100}) a Schr\"odinger-like equation
\begin{equation}\label{450200}
      i\frac{d|\psi(z)\rangle}{dz}=\hat{H}|\psi(z)\rangle,
\end{equation}
subject to the {\it initial condition} $|\psi(z=0)\rangle=\sum_{m=-\infty}^{\infty}\psi_m|m\rangle$.
The "Hamiltonian" of this Schr\"odinger-like equation is given by
\begin{equation}\label{hamil}
     \hat{H} = \omega (-1)^{\hat{n}} + \alpha (\hat{V} + \hat{V}^\dagger ),
\end{equation}
and the linear operators $\hat{V}$ and $\hat{V}^\dagger$ are defined as
\begin{equation}\label{010050}
\hat{V}  \equiv  \sum_{n=-\infty}^{\infty}|n\rangle\langle  n+1 |,\qquad
\hat{V}^{\dagger} \equiv \sum_{n=-\infty}^{\infty}|n+1\rangle\langle  n |,
\end{equation}
acting over the vector space generated by the complete and orthonormal set $\left\{|n\rangle ; n=-\infty,...,+\infty\right\}$. \\
One physical realization of the Hamiltonian (\ref{hamil}) corresponds to the propagation of a quasiparticle among the sites of an infinite one-dimensional chain, when the difference in the site energies at alternate sites is taken to be $2\omega$ and the nearest-neighbor matrix element for site-to-site transfer is $\alpha$ (\cite{kovanis}). Thus, as we already said, the diatomic waveguide array mimics the propagation of quasiparticles. \\
If we propose as solution of the Schr\"{o}dinger-like equation (\ref{450200}) the expansion
\begin{equation}\label{450505}
    |\psi(z)\rangle=\sum_{k=-\infty}^{\infty} u_{n}(z)|n\rangle,
\end{equation}
we obtain for the coefficients $u_{n}(z)=\langle n |\psi(z)\rangle$ the infinite system of first order ordinary differential equations (\ref{450100}). Thus, by solving the Schr\"{o}dinger-like equation \ref{450200} means to solve the system \ref{450100}. \\
The operators $\hat{V}$ and $\hat{V}^\dagger$ are the Suskind-Glogower operators (that are the exponential of the quantum phase operator), and it´s action on the states $|n\rangle$ is
\begin{equation}\label{010060}
    \hat{V}\left\vert n\right\rangle =\left\vert n-1\right\rangle
    \qquad \textrm{and}  \qquad
    \hat{V}^{\dag }\left\vert n\right\rangle =\left\vert n+1\right\rangle.
\end{equation}

\section{Exact solution}
The formal solution to the Schr\"{o}dinger-like equation (\ref{450200}) is
\begin{equation}
   |\psi\rangle =  e^{-i z \hat{H}} |m\rangle,
\end{equation}
where the Hamiltonian $\hat{H}$ is given by (\ref{hamil}) and where we have already used the initial condition $|\psi(0)\rangle=|m \rangle$. \\
Because of the commutation relations of the operators involved in the Hamiltonian is not simple, one can not give a closed form for the  evolution operator.  In order to give a solution, we will make a change of basis, we will go from the discrete basis $ \{|n\rangle; \; n=-\infty,...,\infty\}$ to a continuous basis defined by (the Fourier series)
\begin{equation}\label{451100}
    |\phi\rangle=\sum_{n=-\infty}^{\infty} e^{i n \phi}|n\rangle.
\end{equation}
The inverse transformation is clearly
\begin{equation}\label{451300}
    |n\rangle=\frac{1}{2\pi}\int_{-\pi}^{\pi}d\phi e^{-in\phi}|\phi\rangle.
\end{equation}
Therefore, the solution to the Schr\"{o}dinger-like equation can be written as
\begin{equation}\begin{split}\label{451400}
   |\psi\rangle = \frac{1}{2\pi} \int_{-\pi}^{\pi}d\phi e^{-im\phi} e^{-i z \hat{H}} |\phi\rangle.
\end{split}\end{equation}
Thus, we have to analyze only the last part of the above equation, the action of the operator $e^{-it \hat{H}}$ on $|\phi\rangle$. We write the definition of the exponential operator, we split the series in even and odd powers, we use that
\begin{equation}\begin{split}\label{454100}
    \hat{H}|\phi\rangle= \omega |\phi+\pi \rangle + 2 \alpha \cos\phi  |\phi\rangle,
\end{split}\end{equation}
\begin{equation}\begin{split}\label{454200}
    \hat{H}^2|\phi\rangle=  \Omega^2|\phi\rangle,
\end{split}\end{equation}
and
\begin{equation}\begin{split}\label{454300}
    \hat{H}^2|\phi+\pi \rangle=  \Omega^2|\phi+\pi \rangle,
\end{split}\end{equation}
where
\begin{equation}\label{454305}
    \Omega(\phi)=\sqrt{\omega^2+4\alpha^2 \cos^2\phi}.
\end{equation}
and we get
\begin{equation*}\begin{split}\label{454500}
   |\psi\rangle =\frac{1}{2\pi} \int_{-\pi}^{\pi}d\phi e^{-im\phi} \left[\cos(\Omega z) |\phi\rangle - i 2 \alpha \cos\phi \frac{\sin(\Omega z)}{\Omega}|\phi\rangle -i\omega \frac{\sin(\Omega z)}{\Omega}|\phi+\pi \rangle \right].
\end{split}\end{equation*}
To obtain the solution to the infinite system of differential equations, we recall that by definition $u_n=\langle n |\psi\rangle$ and that $\langle n |\phi\rangle=e^{in\phi}$, so
\begin{eqnarray}\label{454800}
   && u_n (z)= \frac{1}{2\pi }\int _{-\pi }^{\pi }d\phi e^{-i (m-n) \phi }\\ \nonumber&\times&\left\{\cos [\Omega (\phi )z]- i 2 \alpha  \cos \phi \frac{\sin [\Omega (\phi ) z]}{\Omega (\phi )}-(-1)^ni \omega \frac{\sin [\Omega (\phi ) z]}{\Omega (\phi )}\right\}.
\end{eqnarray}
As the waveguide array is symmetric and infinite, we do not loose generality at all if we consider that the "central" waveguide is shined ($m=0$). If the initial state is $|m\rangle=|0\rangle$, and the parity properties of the functions involved are used, the result is reduced to
\begin{equation}\label{455000}
    u_n(z)=\frac{1}{\pi }\int _0^{\pi }\cos (n \phi )\left\{\cos [\Omega (\phi )z] - i \left[2 \alpha  \cos\phi + (-1)^n\omega \right]\frac{\sin [\Omega (\phi ) z]}{\Omega (\phi )}\right\}d\phi
\end{equation}
Note that this solution satisfies the initial conditions. Indeed, if $z=0$ we get $u_n(z)=\frac{1}{\pi }\int _0^{\pi }\cos (n \phi )d\phi$ that it is equal to $\delta_{n,0}$.
Also note that if there is no interaction between the guides; i.e., $\alpha=0$, we obtain
\begin{equation}\label{455100}
    u_n(z)=e^{-i(-1)^n \omega z} \delta_{n,0}
\end{equation}
that it is the solution of the trivial system of equations obtained. \\
In Figure 1, we present the behavior of the exact solution of the guides 0 to 10, with $\omega=1$ and $\alpha=0.3$ for $z$  from $0$ to $100$.  For symmetry reasons the behavior of the guides $-1$ to $-10$ is exactly the same.
\begin{figure}[h!]\label{exacta}
   \centering
   \includegraphics [width=9cm,height=6cm]{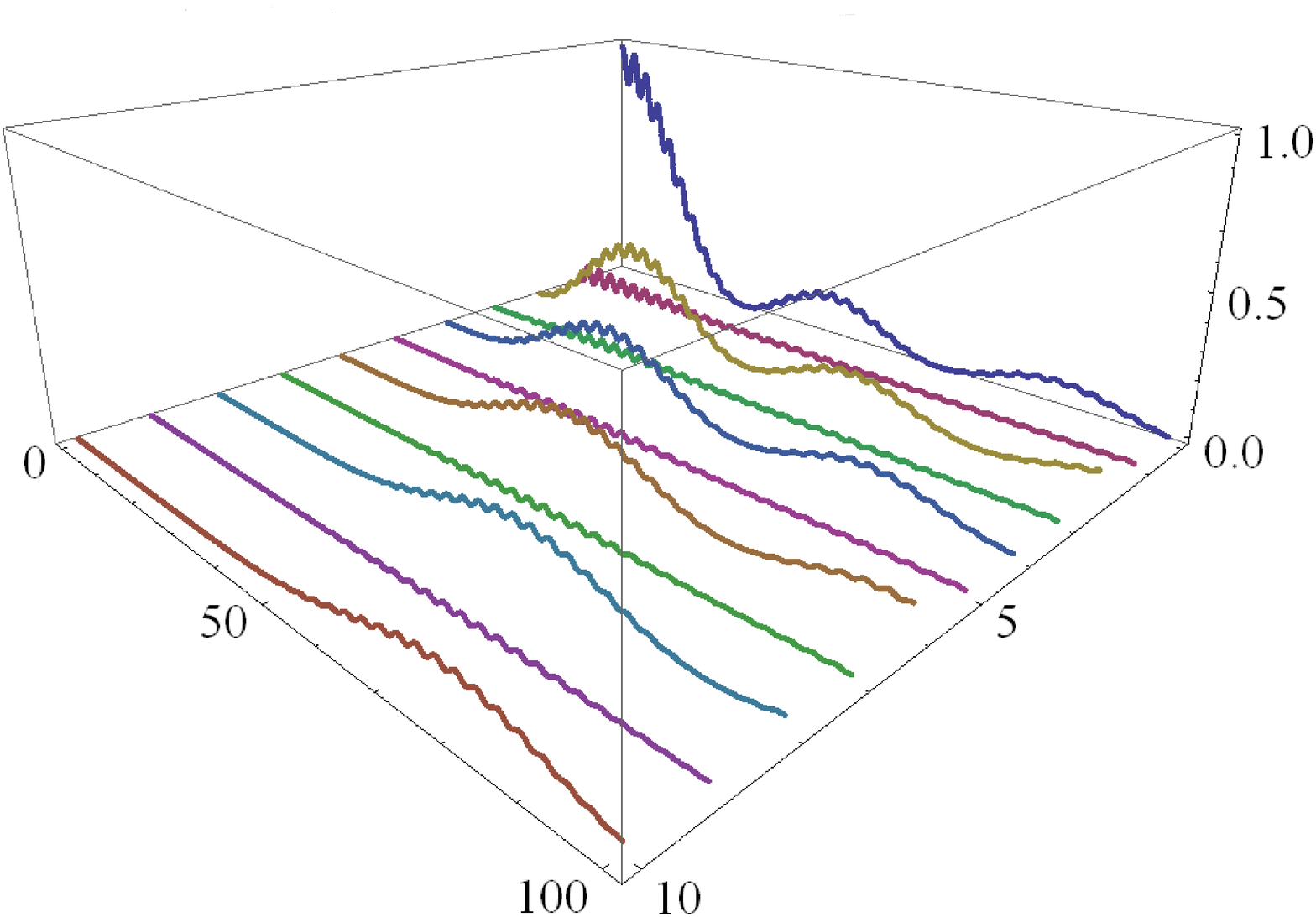}
   \caption{Exact solution of the guides 0 to 10 (for symmetry reasons the behavior of the guides $-1$ to $-10$ is exactly the same) with $\omega=1$ and $\alpha=0.3$ for $z$ from $0$ to $100$}
\end{figure}

\section{Small rotation}
We consider the case in which the parameters obey $\alpha \ll \omega$, and transform the  Hamiltonian
\begin{equation}\label{hamil0}
     \hat{H} = \omega (-1)^{\hat{n}} + \alpha (\hat{V} + \hat{V}^\dagger ),
\end{equation}
via the {\it unitary} transformation
\begin{equation}\label{hamil01}
 \hat{R} =e^{\frac{\alpha}{2\omega} (-1)^{\hat{n}} (\hat{V} + \hat{V}^\dagger )},\qquad \hat{R}^{\dagger}=e^{-\frac{\alpha}{2\omega} (-1)^{\hat{n}} (\hat{V} + \hat{V}^\dagger )},
\end{equation}
such that $ \hat{H}_R=\hat{R} \hat{H} \hat{R}^{\dagger}$. We use the relation
 $e^{s \hat{A}} \hat{B} e^{-s \hat{A}}=\hat{B}+s[\hat{A},\hat{B}]+\frac{s^2}{2!}[\hat{A},[\hat{A},\hat{B}]]+  \dots $
and note that we can cut the series to second order because $\alpha \ll \omega$ \cite{Klimov}, obtaining
\begin{equation}\label{hamil02}
     \hat{H}_R \approx \omega (-1)^{\hat{n}} + \frac{\alpha^2}{2\omega} (-1)^{\hat{n}} (\hat{V} + \hat{V}^\dagger )^2.
\end{equation}
Because the parity operator $ (-1)^{\hat{n}}$ and the operator $(\hat{V} + \hat{V}^\dagger )^2$ commute, we can
obtain the propagator operator, $\hat{U}_R(z)=\exp(-i \hat{H}_R z)$, as the product
\begin{equation}\label{propa}
     \hat{U}_R(z)=e^{-i\omega z(-1)^{\hat{n}}}e^{- i\frac{\alpha^2z}{2\omega} (-1)^{\hat{n}} (\hat{V} + \hat{V}^\dagger )^2}
\end{equation}
and the propagator associated to (\ref{hamil0}) is then
\begin{equation}\label{propa2}
     \hat{U}(z)\approx \hat{R}^{\dagger}e^{-i\omega z(-1)^{\hat{n}}}e^{ - i\frac{\alpha^2z}{2\omega} (-1)^{\hat{n}} (\hat{V} + \hat{V}^\dagger )^2}\hat{R}.
\end{equation}
Expanding the square in the exponential, using once more the commutativity between $(-1)^{\hat{n}}$ and the squared $\hat{V}$ operators and after some trivial algebra, we can write the ket $|\psi(z)\rangle$ as
\begin{equation}\label{20100}
       |\psi (z)\rangle =\hat{R}^{\dagger }e^{-i z \left(\omega +\frac{\alpha ^2}{\omega }\right) (-1)^{\hat{n}} }e^{-i \frac{\alpha ^2}{2\omega } z(-1)^{\hat{n}} \left[\hat{V}^2+(\hat{V}^{\dagger })^2\right]}\hat{R}|\psi (0)\rangle,
\end{equation}
with $ |\psi(0)\rangle$ the initial condition related with the waveguide that is shined.
As $i(-1)^{\hat{n}}(\hat{V}^{\dagger })^2=\left[-i(-1)^{\hat{n}} \hat{V}^2\right]^{-1}$, we may develop the second exponential above in terms of Bessel functions by using their generating function and obtain
\begin{equation}\label{20110}
    |\psi (z)\rangle =\sum _{k=-\infty }^{\infty } (-i)^k J_k\left(\frac{\alpha ^2z}{\omega }\right)\hat{R}^{\dagger }\hat{V}^{2k}e^{-i z \left(\frac{\omega ^2+\alpha ^2}{\omega }\right) (-1)^{\hat{n}} } \left[(-1)^{\hat{n}}\right]^k\hat{R}|\psi (0)\rangle.
\end{equation}
Using that $(-1)^{\hat{n}} \hat{V}^{\dagger }= -\left[(-1)^{\hat{n}}\hat{V}\right]^{-1}$, again the properties of the $V$ operators and the generating function of the Bessel functions, we can write the following explicit expressions for the operators $\hat{R}$ and $\hat{R}^\dag$,
\begin{equation}
    \hat{R}=e^{\frac{\alpha }{2\omega }(-1)^{\hat{n}}\left(\hat{V}+\hat{V}^{\dagger }\right)}=\sum _{j=-\infty }^{\infty } J_j\left(\frac{\alpha }{\omega }\right)\left[ (-1)^{\hat{n}}\hat{V}\right]{}^j,
\end{equation}
and
\begin{equation}
    \hat{R}^{\dagger }=e^{-\frac{\alpha }{2\omega }(-1)^{\hat{n}}\left(\hat{V}+\hat{V}^{\dagger }\right)}=\sum _{\mu =-\infty }^{\infty } (-1)^{\mu }J_{\mu }\left(\frac{\alpha }{\omega }\right)\left[ (-1)^{\hat{n}}\hat{V}\right]{}^{\mu },
\end{equation}
that after being substituted in Equation (\ref{20110}) give us
\begin{equation}\begin{split}\label{20150}
    |\psi (z)\rangle &=\sum _{k=-\infty }^{\infty } \sum _{j=-\infty }^{\infty } \sum _{\mu =-\infty }^{\infty } (-i)^k(-1)^{\mu }J_k\left(\frac{\alpha ^2z}{\omega }\right)J_j\left(\frac{\alpha }{\omega }\right)J_{\mu }\left(\frac{\alpha }{\omega }\right) \\
    &\hat{V}^{2k}\left[ (-1)^{\hat{n}}\hat{V}\right]^{\mu }e^{-i z \left(\frac{\omega ^2+\alpha ^2}{\omega }\right) (-1)^{\hat{n}} }\left[(-1)^{\hat{n}} \right]^k\text{  }\left[ (-1)^{\hat{n}}\hat{V}\right]^j|\psi (0)\rangle.
\end{split}\end{equation}
It is possible to show that
\begin{equation}\label{20160}
    \left[ (-1)^{\hat{n}}\hat{V}\right]^j|m\rangle =(-1)^{j m-\frac{j(j+1)}{2}}|m-j\rangle
\end{equation}
for $j$ positive and negative; so considering the initial condition $|\psi (0)\rangle=|m\rangle$, and after some algebra
\begin{equation}\begin{split}\label{20170}
      u_n(z)=&\langle n|\psi (z)\rangle _m= (-1)^{\frac{m(m-1)-n(n-1)}{2}}\sum _{k=-\infty }^{\infty } \sum _{j=-\infty }^{\infty } (-1)^{k(m-j)} \\
      &e^{-i\text{  }(-1)^{m-j} \left(\frac{\omega ^2+\alpha ^2}{\omega }\right) z }
      i^k J_k\left(\frac{\alpha ^2z}{\omega }\right)J_j\left(\frac{\alpha }{\omega }\right)J_{n-m+2k+j}\left(\frac{\alpha }{\omega }\right)
\end{split}\end{equation}
is obtained. \\
If we consider that the "central" waveguide is shined ($m=0$), we get
\begin{equation}\begin{split}\label{20180}
   u_n(z)=&(-1)^{\frac{n(n-1)}{2}}\sum _{k=-\infty }^{\infty } \sum _{j=-\infty }^{\infty } (-1)^{j k}e^{-i\text{  }(-1)^j\left(\frac{\omega ^2+\alpha ^2}{\omega }\right)z\text{  }} \\
   &i^kJ_k\left(\frac{\alpha ^2}{\omega }z\right)J_j\left(\frac{\alpha }{\omega }\right)J_{n+2k+j}\left(\frac{\alpha }{\omega }\right),
\end{split}\end{equation}
that is the final solution in this approximation. \\
Using that
\begin{equation}\label{20190}
    \sum _{j=1}^{\infty } \left.[J_j\left(x\right)\right]{}^2=\frac{1}{2}\left\{1-\left[J_0\left(x\right)\right]{}^2\right\},
\end{equation}
and that
\begin{equation}\label{20200}
    \sum _{j=-\infty }^{\infty } J_j\left(x\right)J_{n+j}\left(x\right)=0,
\end{equation}
it is possible to show that the initial conditions are satisfied. It is also easy to verify that in the special case when $\alpha=0$, we get the correct very well known trivial solution.
\begin{figure}[h!]\label{sr0.1}
   \centering
   \includegraphics [width=12cm,height=6cm]{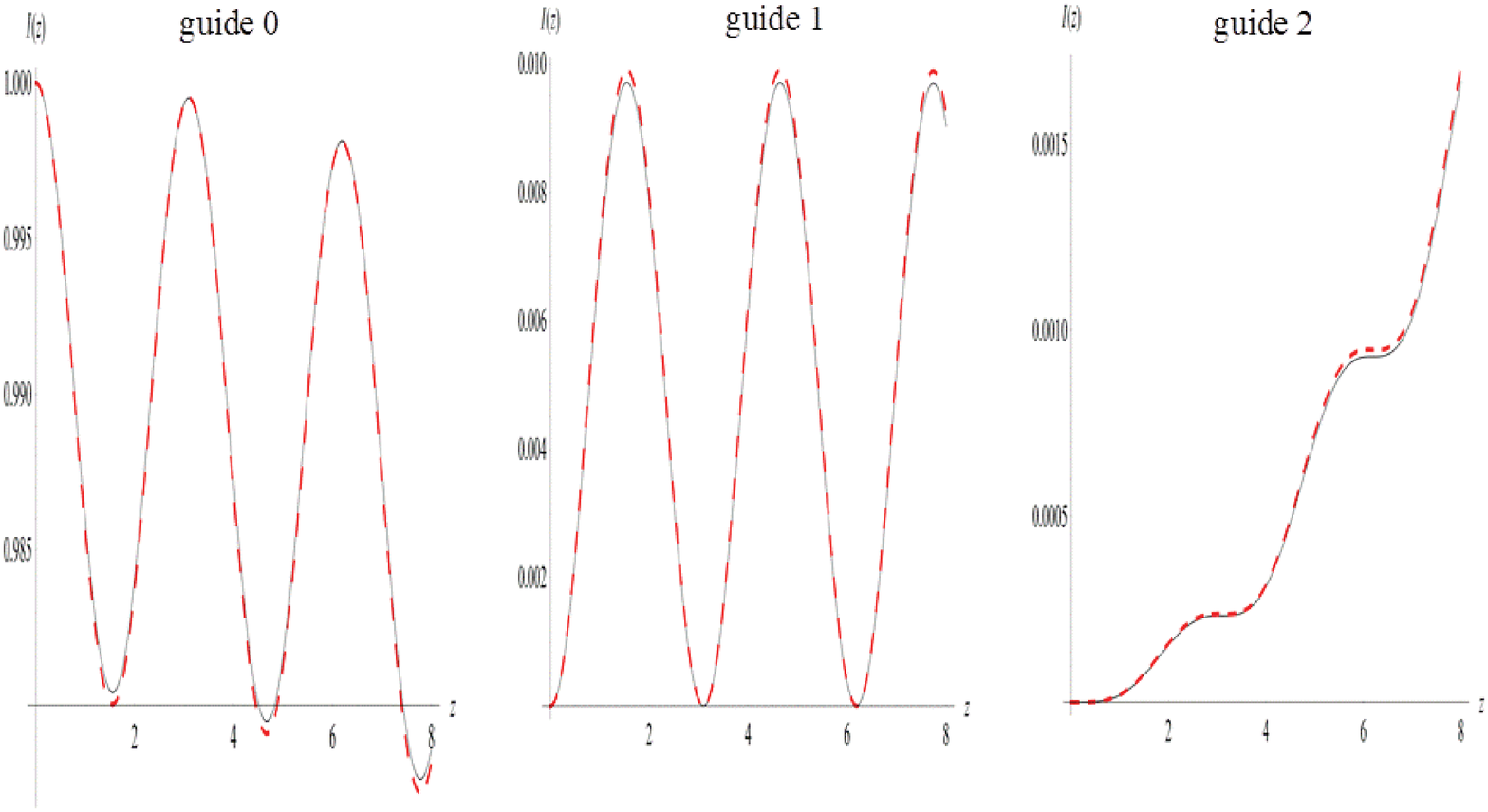}
   \caption{Comparison between the exact numerical solution (continuous line) and the small rotation approximation solution (dashed line) for $\omega=1$ and $\alpha=0.1$, for the first three guides}
\end{figure}
\begin{figure}[h!]\label{sr0.3}
   \centering
   \includegraphics [width=12cm,height=6cm]{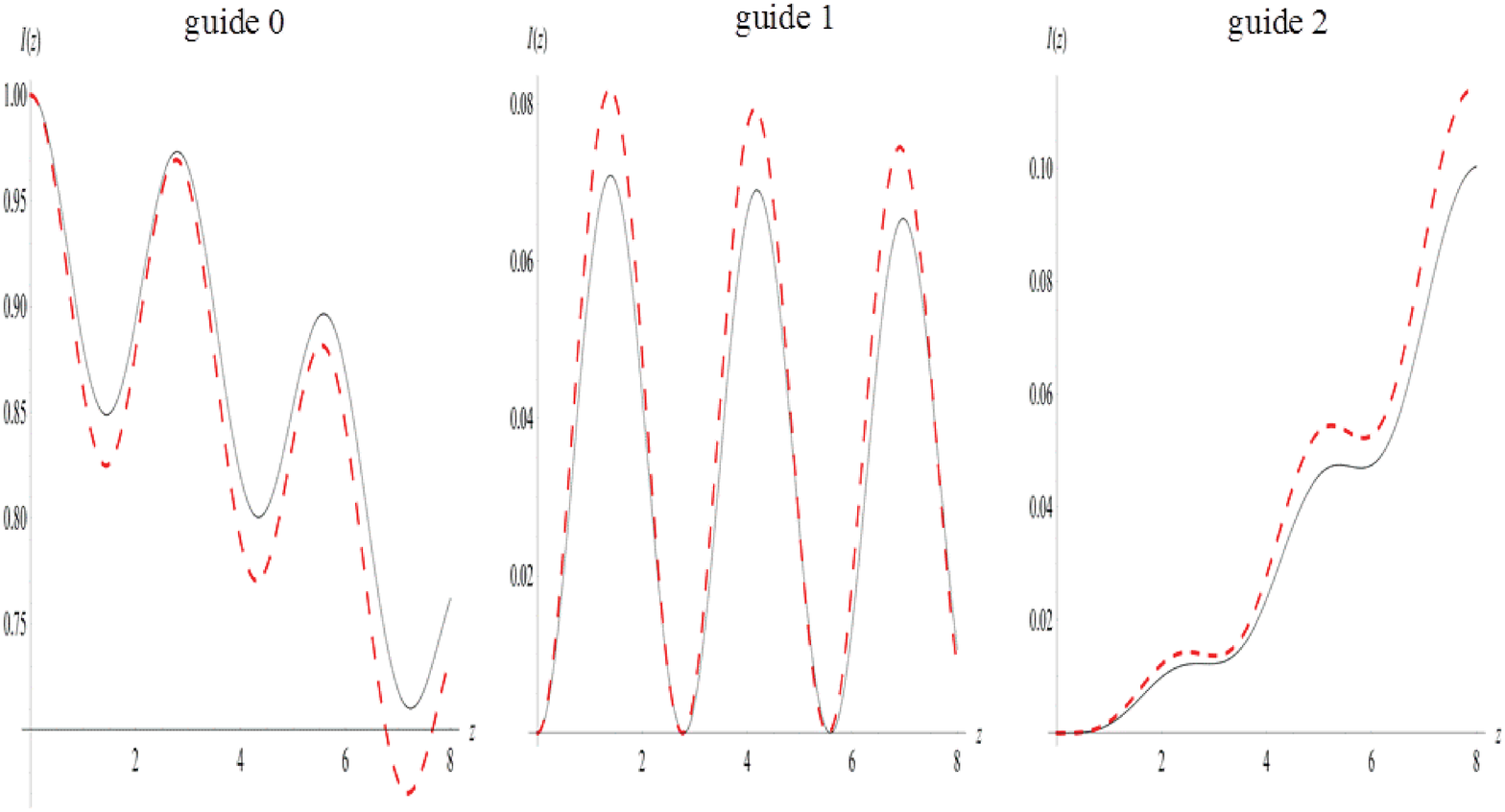}
   \caption{Comparison between the exact numerical solution (continuous line) and the small rotation approximation solution (dashed line) for $\omega=1$ and $\alpha=0.3$, for the first three guides}
\end{figure}
In Figure 2, we compare the exact solution and the small rotation approximation for $\alpha=0.1$ and in Figure 3 for $\alpha=0.3$ giving still good accuracy even for larger $\alpha$'s. \\
The small rotation approach is similar to adiabatic elimination in the case of atoms interacting with quantized light, where, effectively atomic levels may be eliminated by a proper rotation. In our case, it has the same meaning as the diatomic waveguide array may be viewed as a two level system \cite{Longhi}, and we effectively obtain a one level system, which is integrable.\\

It must be remarked that a traditional Rayleigh-Schr\"{o}dinger pertubative solution of equation (\ref{450200}) can be investigated. Considering $\omega (-1)^{\hat{n}}$ as the non-perturbed part, $\hat{V} + \hat{V}^\dagger$ as the perturbation, and $\alpha$ as the "smallness" parameter, the solution obtained to the initial system of differential equations (\ref{450100}) is
\begin{equation}\begin{split}
    u_n(z)=&e^{-(-1)^mi \omega  z}\delta _{n,m}+\sum _{k=1}^{\infty } \alpha^k \frac{(-1)^{m k}}
    {2^k\left[\frac{k}{2}\right]!\omega ^k}\{e^{-i z \omega (-1)^{m+k}}P_k\left[2i z \omega (-1)^{m+k}\right] \\
    &+e^{i z \omega (-1)^{m+k}}Q_k\left[2i z \omega (-1)^{m+k}\right]\}\sum _{j=0}^k \left(\begin{array}{c} k \\ j\end{array}\right)\delta _{n,m+k-2j}
\end{split}\end{equation}
where $[k]$ is the entire part of $k$, and $P_j(\xi)$ and $Q_m(\xi)$ are polynomials. The two families of polynomials satisfies the recurrence relations
\begin{equation}\label{pol1}
    R_{2j}(\xi)=(2j-1)R_{2j-1}(\xi)-\xi R_{2j-2}(\xi) , \qquad  j=1,2,3, \cdots
\end{equation}
and
\begin{equation}\label{pol2}
    R_{2j+1}(\xi)=-2R_{2j-1}(\xi)+\xi R_{2j-1}(\xi) , \qquad  j=1,2,3, \cdots
\end{equation}
where $R$, stands for $P$ or $Q$. The "\textit{P}" family starts with the initial polynomials
\begin{equation}\label{pol3}
    P_0(\xi)=1 , \qquad  P_1(\xi)=-1
\end{equation}
and the "\textit{Q}" family with
\begin{equation}\label{pol4}
    Q_0(\xi)=0 , \qquad  Q_1(\xi)=1.
\end{equation}
\begin{figure}[h!]
   \centering
   \includegraphics [width=12cm,height=6cm]{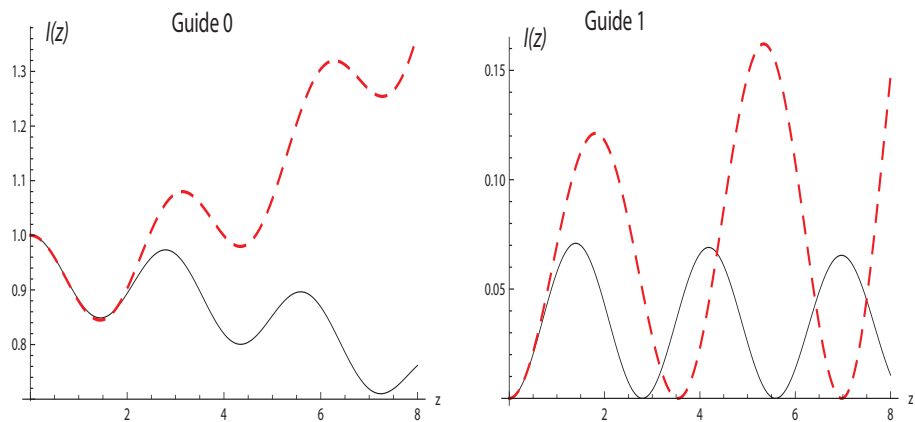}
   \caption{Comparison between the exact numerical solution (continuous line) and the Rayleigh-Schr\"{o}dinger perturbative solution to third order (dashed line) for $\omega=1$ and $\alpha=0.3$, for the first two guides.}
\end{figure}
However, as it is shown in Figure 4, this approximated solution diverges with time. The Rayleigh-Schr\"odinger perturbation approach had already been proposed previously by Rother \cite{Rother}, and he also found convergence problems. To overcome this convergence problem the small rotation approach is proposed.

\section{Conclusions}
As the traditional Rayleigh-Schr\"odinger perturbation approach to the Schr\"{o}dinger type equation (\ref{450200}) gives rise to convergence problems, a small rotation approximation method is applied; the convergency problems are eliminated, and in the small rotation situation we can obtain, to a good degree of accuracy, approximated solutions to the problem of light propagating in waveguide arrays. \\
The small rotation assumption used corresponds, in the case of the propagation of a quasiparticle among the sites of an infinite one-dimensional chain, to suppose that the nearest-neighbor matrix element for site-to-site transfer is small.


\begin{thebibliography}{99}

\bibitem{Crisp} Crisp M D 1993 Ed Jaynes' steak dinner problem II ({\it Physics and Probability, Essays in honor of Edwin T. Jaynes}) ed. W.T. Grandy, Jr, and P.W. Milonni (Cambridge: Cambridge University Press).

\bibitem{Jona}  Moya-Cessa H, Jonathan D and  Knight P L, 2003 {\it J. of Mod. Optics} {\bf 50}, 265-273.

\bibitem{Champi} Moya-Cessa H, Soto-Eguibar F, Vargas-Martinez J M, Juarez-Amaro R, Zuniga-Segundo A, 2012 {\it Physics Reports} {\bf 512}, 229-261.

\bibitem{Manko1} Man'ko M A, Man'ko V I and Vilela Mendes R 2001 {\it Phys. Lett.} {\bf 288} 132--138.

\bibitem{Longhi} Longhi S 2011 {\it Opt. Lett.} {\bf 36}  3407--3409.

\bibitem{Longhi2} Crespi A, Longhi S and Osellame R 2012 {\it Phys. Rev. Lett.} {\bf 108} 163601.

\bibitem{Manko2} Man'ko V I, Marmo G, Sudarshan E C G and Zaccaria F 1997 {\it Phys. Scripta} {\bf 55} 528--541.

\bibitem{Roberto} Le\'on-Montiel R de J  and Moya-Cessa H 2011 {\it Int. J. of Quant. Inf.} {\bf 9} 349--355.

\bibitem{Keil} Keil R, Perez-Leija A, Dreisow F, Heinrich M, Moya-Cessa H, Nolte S, Christodoulides D N, and Szameit A 2011 {\it Phys. Rev. Lett.} {\bf 107} 103601.

\bibitem{kovanis} Kovanis V.I. and Kenkre V.M. (1988) Physics Letters 130,3, 147.

\bibitem{perez} Perez-Leija A, Moya-Cessa H, Szameit A, and Christodoulides D N 2010 {\it Opt. Lett.} {\bf 35} 2409--2411.

\bibitem{Klimov} Klimov A and  S\'anchez-Soto L L 2000 {\it Phys. Rev.} A, {\bf 61}, 063802.

\bibitem{Rother} Rother T 1993, Journal of Electromagnetic Waves and Applications 7, 857.

\end{thebibliography}
\end{document}